\documentstyle[preprint,aps,eqsecnum,floats]{revtex}
\def\abstracts#1{{
     \tenrm\baselineskip=12pt\noindent
        \centerline{\tenrm ABSTRACT}\vspace{0.3cm}
        \parindent=0pt #1
         \par}}

\begin{document}
\baselineskip=12pt
\vspace{1.9cm}
\title{\baselineskip=18pt
COMMENSURATE SCALE RELATIONS: PRECISE TESTS OF
       QUANTUM CHROMODYNAMICS WITHOUT SCALE OR SCHEME AMBIGUITY
\thanks{\baselineskip=14pt
Invited talk given by Stanley J. Brodsky at the
Tennessee International Symposium on Radiative
Corrections: Status and Outlook,
June 27--July 1, 1994,  Gatlinburg, Tennessee.
Work partially supported by the Department of
Energy, contract DE--AC03--76SF00515 and contract
DE--FG02--93ER--40762.}}
\author{STANLEY J. BRODSKY}
\vspace{.5cm}
\address{\baselineskip=14pt
Stanford Linear Accelerator Center \\
Stanford University, Stanford, California 94309}
\author{HUNG JUNG LU}
\vspace{.5cm}
\address{\baselineskip=14pt
Department of Physics, University of Maryland\\
College Park, Maryland 20742}
\maketitle
\vspace{0.9cm}
\abstracts{
We derive commensurate scale relations which relate
perturbatively calculable QCD
observables to each other, including the annihilation ratio $R_{e^+e^-},$
the heavy quark potential, $\tau$ decay, and  radiative
corrections to structure function
sum rules.  For each such observable one can define
an effective charge, such as
$\alpha_R(\sqrt s)/\pi \equiv R_{e^+e^-}(\sqrt s)/(3 \sum e^2_q) -1.$
The commensurate scale relation connecting the effective charges for
observables $A$ and $B$ has the form
$\alpha_A(Q_A) = \alpha_B(Q_B)
 \left(1 + r_{A/B} {\alpha_B\over \pi} +\cdots\right),$
where the coefficient $r_{A/B}$ is independent of the number of flavors
$f$ contributing to coupling renormalization, as in BLM scale-fixing.
The ratio of scales $Q_A/Q_B$ is unique at leading order and guarantees
that the observables $A$ and $B$ pass through new quark thresholds at the
same physical scale.  In higher orders a different renormalization scale
$Q^{n*}$ is assigned for each order $n$ in the perturbative series such
that the coefficients of the series are identical to that of a
conformally invariant theory.  The commensurate scale relations and
scales satisfy the renormalization group transitivity rule which ensures
that predictions in PQCD are independent of the choice of an intermediate
renormalization scheme $C.$ In particular, scale-fixed predictions can be
made without reference to theoretically constructed singular
renormalization schemes such as $\overline{\rm MS}.$ QCD can thus be
tested in a new and precise way by checking that the effective charges of
observables track both in their relative normalization and in their
commensurate scale dependence.  The commensurate scale relations which
relate the radiative corrections to the annihilation ratio $R_{e^+e^-}$
to the radiative corrections for the Bjorken and Gross-Llewellyn Smith
sum rules are particularly elegant and interesting.  The final series has
simple coefficients which are independent of color:
$\widehat\alpha_{g_1}(Q) = \widehat\alpha_R(Q^*) -
\widehat\alpha_R^2(Q^{**}) + \widehat\alpha_R^3(Q^{***})+\cdots$, where
$\widehat\alpha = ({3 C_F / 4 \pi}) \alpha.$
The coefficients coincide with the Crewther relation obtained in
conformally invariant gauge theory. Thus this commensurate scale
relation provides the generalization of the Crewther relation
to non-conformal gauge theory.}
\thispagestyle{empty}
\newpage

\baselineskip 12pt
\vskip1cm
\leftline{{\bf 1. Introduction}}
\vskip.5cm

The problem of the renormalization scale dependence of perturbative QCD
predictions has plagued attempts to make reliable and precise tests of
the theory.  There is, in fact, no consensus on how to estimate the
theoretical error due to the scale ambiguity, what constitutes a
reasonable range of physical values, or indeed how to identify what the
central value should be.  The problem is compounded in multi-scale
problems where several plausible physical scales enter.  Even worse, if
we consider the renormalization scale as totally arbitrary, the
next-to-leading coefficient in the perturbative expansion can take on the
value zero or any other value.  Thus it is difficult to assess the
convergence of the truncated series, and finite-order analyses cannot be
meaningfully compared to experiment.

Recently we have shown how the scale ambiguity problem can be avoided by
focussing on relations between experimentally measurable
observables \cite{BrodskyLuBig}. The conventional $\overline{\rm MS}$
renormalization scheme serves simply as an intermediary between
observables. For example, consider the entire radiative corrections to
the annihilation cross section expressed as the ``effective charge"
$\alpha_R(Q)$ where $Q=\sqrt s$:
\begin{equation}
 R(Q) \equiv 3 \sum_f Q_f^2 \left[ 1+
{\alpha_R(Q) \over \pi} \right].
\end{equation}
Similarly, we can define the entire radiative correction to the Bjorken
sum rule as the effective charge $\alpha_{g_1}(Q)$ where $Q$ is the
lepton momentum transfer:\footnote{\baselineskip 12pt Here we follow the
normalization given in Ref. \cite{LarinVermaseren}.  For a recent review
on this sum rule, see Ref. \cite{EllisKarliner}.}
\begin{equation}
\int_0^1
d x \left[
   g_1^{ep}(x,Q^2) - g_1^{en}(x,Q^2) \right]
   \equiv {1\over 3} \left|g_A \over g_V \right|
   \left[ 1- {\alpha_{g_1}(Q) \over \pi} \right] .
\end{equation}

We now use the known expressions to three loops in $\overline{\rm MS}$
scheme and choose the scales $Q^*$ and $Q^{**}$ to re-sum all quark and
gluon vacuum polarization corrections into the running couplings.  The
values of these scales are the physical values of the energies or
momentum transfers which ensure that the radiative corrections to each
observable passes through the heavy quark thresholds at their respective
commensurate physical scales.  The final result is remarkably simple:
\begin{equation}
{\alpha_{g_1}(Q) \over \pi} = {\alpha_R(Q^*) \over \pi} -
\left( {\alpha_R(Q^{**}) \over \pi} \right)^2 + \left( {\alpha_R(Q^{***})
\over \pi} \right)^3 + \cdots .
\label{AlphaG1AlphaRAfterBLMThreeFlavors}
\end{equation}
The coefficients in the series (aside for a factor of $C_F,$ which can be
absorbed in the definition of $\alpha_s$) are actually independent of
color and are the same in Abelian, non-Abelian, and conformal gauge
theory.  The non-Abelian structure of the theory is reflected in the
scales $Q^*$ and $Q^{**}.$

Any perturbatively calculable physical quantity can be used to define an
effective charge \cite{Grunberg,DharGupta,GuptaShirkovTarasov} by
incorporating the entire radiative correction into its definition.  An
important result is that all effective charges $\alpha_A(Q)$ satisfy the
Gell-Mann-Low renormalization group equation with the same $\beta_0$ and
$\beta_1;$ different schemes or effective charges only differ through the
third and higher coefficients of the $\beta$ function.  Thus, any
effective charge can be used as a reference running coupling constant in
QCD to define the renormalization procedure.  More generally, each
effective charge or renormalization scheme, including $\overline{\rm
MS}$, is a special case of the universal coupling function $\alpha(Q,
\beta_n)$ \cite{BrodskyLuStueckelbergPeterman}.  Peterman and
St\"uckelberg have shown \cite{StueckelbergPeterman} that all effective
charges are related to each other through a set of evolution equations in
the scheme parameters $\beta_n.$

We shall refer to the connections between the effective charges of
observables such as Eq. (\ref{AlphaG1AlphaRAfterBLMThreeFlavors}) as
``commensurate scale relations" (CSR) \cite{BrodskyLuBig}.  A fundamental
test of QCD will be to verify empirically that the observables track in
both normalization and shape as given by the CSR.  The commensurate scale
relations thus provide fundamental tests of QCD which can be made
increasingly precise and independent of the choice of renormalization
scheme or other theoretical convention.

The relation between physical observables must be independent of the
choice of any intermediate renormalization scheme.  In a renormalizable
quantum field theory, such as quantum chromodynamics, the Lagrangian
${\cal L}[g(\mu), m_i(\mu), ...]$ is written a function of the coupling
$g(\mu)$ and masses $m_i(\mu)$ at a given scale $\mu.$ In principle, one
can choose a minimal set of input measurements to fix these bare
parameters, and then predict all other physical observables.  For
example, in quantum electrodynamics, it is conventional to use Coulomb
scattering $\ell \ell \to \ell \ell$ extrapolated to zero momentum
transfer and threshold Compton scattering $\gamma \ell \to \gamma \ell$
to fix the bare charge and lepton mass parameters.  One then can
systematically predict the values of the lepton anomalous magnetic
moments and the other high precision QED observables.  Renormalization
group equations reflect the fact that predictions connecting the input
and output observables can be computed without ambiguity order by order
in perturbation theory; {\it i.e.}, the relations connecting physical
observables are independent of the choice of the scale $\mu$ and the
choice of intermediate renormalization scheme.

The theoretical predictions connecting any pair of perturbatively
calculable observables are given explicitly by the commensurate scale
relations.  In these relations between the effective charges of the
physical observables, the dependence on the value of the bare coupling
and the choice of renormalization procedure cancels out, as expected,
order by order in perturbation theory.  Thus the CSR provide precise and
direct experimental tests of quantum field theory without scale or scheme
ambiguity.

The CSR for observables $A$ and $B$ in terms of their effective charges
has the form
\begin{equation}
\alpha_A(Q_A) = \alpha_B(Q_B)\ \left(1 + r_{A/B} {\alpha_B\over
\pi} +\cdots\right)\ .
\label{BLMEq}
\end{equation}
The ratio of the scales $\lambda_{A/B}=Q_A/Q_B$ is fixed by the
requirement that the coefficient $r_{A/B}$ is independent of the number
of flavors $f$ contributing to coupling constant renormalization.  This
guarantees that the effective charges for the observables $A$ and $B$
pass through new quark thresholds at the same physical scale.  The scales
$Q_A$ and $Q_B$ are thus commensurate.  The value of $\lambda_{A/B}$ is
unique at leading order, and the relative scales satisfy the transitivity
rule\cite{BrodskyLuSelfconsistency}
\begin{equation}
\lambda_{A/B} = \lambda_{A/C}~\lambda_{C/B}\ .
\end{equation}
This is equivalent to the group property defined by Peterman and
St\"uckelberg \cite{StueckelbergPeterman} which ensures that predictions
in PQCD are independent of the choice of an intermediate renormalization
scheme $C$ \footnote{We thank A.  Kataev for an illuminating discussion
on this point.} (The renormalization group method was developed by
Gell-Mann and Low \cite{GellMannLow} and by Bogoliubov and Shirkov
\cite{BogoliubovShirkov}.) In particular, scale-fixed predictions can be
made without reference to theoretically constructed renormalization
schemes such as $\overline{\rm MS};$ QCD can thus be tested by checking
that the observables track both in their relative normalization and
commensurate scale dependence \cite{Leipzig}.

The transitivity and symmetry properties of the commensurate scales are
the scale transformations of the renormalization ``group" as originally
defined by Peterman and St\"uckelberg \cite{StueckelbergPeterman}.  The
predicted relation between observables must be independent of the order
one makes substitutions; {\it i.e.} the algebraic path one takes to
relate the observables.  Furthermore, any method which fixes the scale in
QCD must also be applicable to Abelian theories such as QED, since in the
limit of small number of colors $N_C \to 0$ or large number of flavors
$f$ the perturbative coefficients in QCD coincide with the perturbative
coefficients of an Abelian analog of QCD\footnote{We thank Patrick Huet
and Eric Sather for conversations on this point.}.

The relation between scales in the CSR is consistent with the BLM
scale-fixing procedure\cite{BrodskyLepageMackenzie} in which the scale is
chosen such that all terms arising from the QCD $\beta-$function are
resummed into the coupling.  Note that this also implies that the
coefficients in the perturbation CSR expansions are independent of the
number of quark flavors $f$ renormalizing the gluon propagators.  This
prescription ensures that, as in quantum electrodynamics, vacuum
polarization contributions due to fermion pairs are all incorporated into
the coupling $\alpha(\mu)$ rather than the coefficients.

One of the most useful observables in QCD is the heavy quark potential,
since it can be computed in lattice gauge theory from a Wilson loop.
Since interacting quarks have infinite mass, the radiative correction are
associated with the exchange diagrams, rather than the vertex
corrections.  It is convenient to write the heavy quark potential as
$V(Q^2)=-4\pi C_F \alpha_V(Q)/Q^2.$ This defines the effective charge
$\alpha_V(Q^2)$ where by definition the ``self-scale" $Q^2= -t$ is the
momentum transfer squared.

The BLM scale has also recently been used by Lepage and Mackenzie
\cite{LepageMackenzie} and their co-workers to improve lattice
perturbation theory.  By using the BLM method one can eliminate
$\alpha_{\rm Lattice}$ in favor of $\alpha_V$ thus avoiding an expansion
with artificially large coefficients.  An essential step in this
derivation is the use of the BLM method to connect the scale $\pi / a$ in
lattice perturbation theory to the physical scale appearing in
$\alpha_V$.  The lattice determination, together with the empirical
constraints from the heavy quarkonium spectra, promises to provide a
well-determined effective charge $\alpha_V(Q)$ which could be adopted as
the QCD standard coupling.  In fact, a precise determination
$\alpha_V(8.2\ {\rm GeV})= 0.1957(34)$ has recently been obtained
(including 3 dynamical fermions) from
the $\Upsilon$ spectrum by the Cornell lattice group \cite{Davies}.

In the case of non-Abelian theory, the BLM method automatically resums
the corresponding gluon as well as quark vacuum polarization
contributions since the coupling $\alpha_s$ is a function of $\beta_0
\propto 11-{2\over 3} f.$ For example in the $\alpha_V$ scheme the
coupling is defined to sum all vacuum polarization contribution, so that
coefficients of the expansion in powers of this coupling cannot depend on
the number of flavors $f$ arising from vacuum polarization.  The
transitivity property of the renormalization group equations
then requires that this is true when expanding in any effective charge.

\begin{center}
\setlength{\unitlength}{1in}
\begin{picture}(4,2.8)

\put(1.7,2.5){$\underline{\rm Table \ I}$}

\put(0.5,2.2){Leading Order Commensurate Scale Relations}

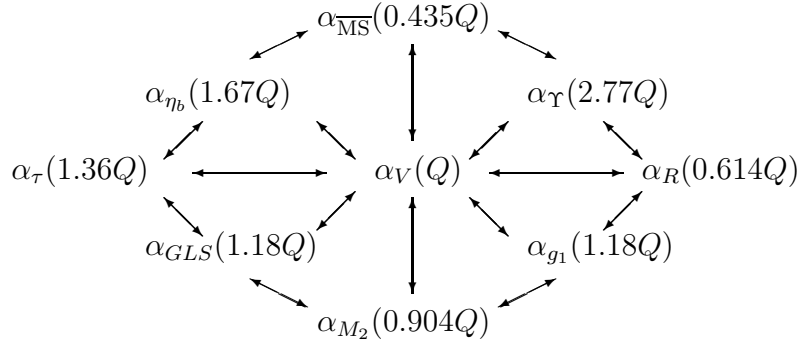
\put(1.6,1.7){$\alpha_{\overline{\rm MS}}(0.435 Q)$}
\put(0.7,1.3){$\alpha_{\eta_b }(1.67 Q)$}
\put(2.7,1.3){$\alpha_\Upsilon (2.77 Q)$}
\put(0.0,0.9){$\alpha_\tau(1.36 Q)$}
\put(1.9,0.9){$\alpha_V(Q)$}
\put(3.3,0.9){$\alpha_R(0.614 Q)$}
\put(0.7,0.5){$\alpha_{GLS}(1.18 Q)$}
\put(2.7,0.5){$\alpha_{g_1}(1.18 Q)$}
\put(1.6,0.1){$\alpha_{M_2}(0.904 Q)$}

\put(1.4,1.6){\vector(2,1){0.15}}
\put(1.4,1.6){\vector(-2,-1){0.15}}
\put(2.7,1.6){\vector(-2,1){0.15}}
\put(2.7,1.6){\vector(2,-1){0.15}}
\put(1.4,0.3){\vector(-2,1){0.15}}
\put(1.4,0.3){\vector(2,-1){0.15}}
\put(2.7,0.3){\vector(2,1){0.15}}
\put(2.7,0.3){\vector(-2,-1){0.15}}
\put(0.8,1.0){\vector(1,1){0.2}}
\put(1.0,1.2){\vector(-1,-1){0.2}}
\put(3.3,1.0){\vector(-1,1){0.2}}
\put(3.1,1.2){\vector(1,-1){0.2}}
\put(1.0,0.6){\vector(-1,1){0.2}}
\put(0.8,0.8){\vector(1,-1){0.2}}
\put(3.1,0.63){\vector(1,1){0.2}}
\put(3.3,0.83){\vector(-1,-1){0.2}}

\put(2.4,1.0){\vector(1,1){0.2}}
\put(2.6,1.2){\vector(-1,-1){0.2}}
\put(1.8,1.0){\vector(-1,1){0.2}}
\put(1.6,1.2){\vector(1,-1){0.2}}
\put(2.6,0.6){\vector(-1,1){0.2}}
\put(2.4,0.8){\vector(1,-1){0.2}}
\put(1.6,0.63){\vector(1,1){0.2}}
\put(1.8,0.83){\vector(-1,-1){0.2}}

\put(2.1,1.35){\vector(0,1){0.25}}
\put(2.1,1.35){\vector(0,-1){0.25}}
\put(2.1,0.55){\vector(0,1){0.25}}
\put(2.1,0.55){\vector(0,-1){0.25}}

\put(1.3,0.93){\vector(1,0){0.35}}
\put(1.3,0.93){\vector(-1,0){0.35}}
\put(2.85,0.93){\vector(1,0){0.35}}
\put(2.85,0.93){\vector(-1,0){0.35}}

\end{picture}
\end{center}

\noindent
Alternatively, one can derive the leading-order BLM scale when expanding
in $\alpha_V$ by explicitly integrating the one loop integrals in the
calculation of the observable $A$ using $\alpha_V(\ell^2)$ in the
integrand, where $\ell^2$ is the four-momentum transferred squared
carried by the gluon. (In practice one only needs to compute the
mean-value $\langle \ell n \ \ell^2 \rangle = \ell n\, Q^2_V
$\cite{LepageMackenzie}.)

In general, the coefficients in the perturbative expansion using BLM
scale fixing are the same as those of the corresponding conformally
invariant theory with $\beta=0.$ In practice, the conformal limit is
defined by $\beta_0, \beta_1 \to 0$, and can be reached, for instance, by
adding enough spin-half and scalar quarks as in supersymmetric QCD with
$N_C=4$.  Since all the running coupling effects have been absorbed into
the renormalization scales, the scale setting method described here
correctly reproduces the expansion coefficients in this
limit.\footnote{We thank Dieter M\"uller for discussions on this point.}
It should be pointed out that other scale-setting procedures do not
guarantee this feature.

The scale-fixed relation between the heavy quark potential effective
charge and $\alpha_{\overline{\rm MS}}$ is
$\alpha_V(Q)=\alpha_{\overline{\rm MS}}\,(e^{-5/6}Q)[1 - 2
(\alpha_{\overline{\rm MS}}/\pi) + \cdots ]$
\cite{BrodskyLepageMackenzie}. (The one-loop calculation of $\alpha_V$ in
$\overline{\rm MS}$ scheme is given in Ref. \cite{HeavyQuarkPotential}.)
The new result of Ref. \cite{Davies} then implies
$\alpha^{(3)}_{\overline {\rm MS}}(3.56\ {\rm GeV}) = 0.2201(84)$ and a
very precise value at $Q=M_z$: $\alpha^{(5)}_{\overline{\rm
MS}}(M_z)=0.115(2)$.  This commensurate scale relation provides an
analytic definition of $\overline{\rm MS}$ scheme which automatically
incorporates the transition between quark flavors.

The physical value of the commensurate scale in $\alpha_V$ scheme
reflects the mean virtuality of the exchanged gluon.  However, in other
schemes, including $\overline{\rm MS},$ the argument of the effective
charge is displaced from its physical value.  The relative scale for a
number of observables is indicated in Table I.  For example, the physical
scale for the branching ratio $\Upsilon \to \gamma X$ when expanded in
terms of $\alpha_V$ is $(1/2.77) M_\Upsilon \sim (1/3) M_\Upsilon,$ which
reflects the fact that the final state phase space is divided among three
vector systems. (When one expands in $\overline{\rm MS}$ scheme, the
corresponding scale is $0.157 M_\Upsilon.$) Similarly, the physical scale
appropriate to the hadronic decays of the $\eta_b$ is $(1/1.67)
M_{\eta_b} \sim (1/2) M_{\eta_b}.$

The BLM method has recently been applied to the analysis of jet ratios
in $e^+ e^-$ collisions by Kramer and Lampe\cite {KramerLampe} and jet
ratios in $e p$ collisions by Ingelman and Rathsman
\cite{IngelmanRathsman}.  One can determine the scale $Q^*$ for
$(2+1)$ jets at HERA as a function of all of the available scales.
The method has also been applied to the radiative corrections to the top
width decay by Voloshin and Smith \cite {VoloshinSmith} and to other
electroweak measures by Sirlin \cite {Sirlin}.

\vskip1cm
\leftline{{\bf 2. CSR in Higher Order}}
\vskip.5cm

It is straightforward to derive the general connection between any two
observables in QCD through order $\alpha_s^3$ if one is given the
calculation of each of the observables to NNLO in $\overline{\rm MS}$
scheme.  As we have emphasized, the choice of the intermediate scheme is
irrelevant for the final CSR connecting the two observables.

The first step is to eliminate the intermediate scheme algebraically.
The expansion series of a physical effective charge $\alpha_1(Q)/\pi$ in
terms of another physical effective charge $\alpha_2(Q)/\pi$ has the
form\cite{BrodskyLuBig}
\begin{eqnarray}
\frac{\alpha_1(Q)}
     {\pi}
&=&
\frac{\alpha_2(Q)}
     {\pi}
+
(A_{12}+B_{12} f)
\left(
       \frac{\alpha_2(Q)}
            {\pi}
\right)^2
\nonumber
\\
&\quad +& \
(C_{12} + D_{12} f+ E_{12} f^2)
\left(
       \frac{\alpha_2(Q)}
            {\pi}
\right)^3
+ \cdots .
\label{Alpha1Alpha2BeforeBLM}
\end{eqnarray}
After the resummation of running coupling effects into a set of new
renormalization scales $Q^*$, $Q^{**}$ and $Q^{***}$., one arrives at a
series of the form\footnote{\baselineskip 12pt We must exclude from the
analysis the potential $f$ dependence in the NNLO term induced by
light-by-light diagrams.  These diagrams are finite and do not
participate in the renormalization of the running coupling constant.  As
a convention, the coefficient $D_{12}$ in Eq.
(\ref{Alpha1Alpha2BeforeBLM}) will include only the $f$ dependence from
the running of the coupling constant.  The extra $f$-dependent terms from
light-by-light scattering diagrams will be considered as part of the
$C_{12}$.  This separation is straightforward in the practical examples
considered here.}
\begin{eqnarray}
\frac{\alpha_1(Q)}
     {\pi}
&=&
\frac{\alpha_2(Q^*)}
     {\pi}
+
{\widetilde A}_{12}
\left(
       \frac{\alpha_2(Q^{**})}
            {\pi}
\right)^2
+
{\widetilde C}_{12}
\left(
       \frac{\alpha_2(Q^{***})}
            {\pi}
\right)^3
+ \cdots .
\label{Alpha1Alpha2AfterBLM}
\end{eqnarray}
\begin{eqnarray}
\widetilde A_{12}
&=&
A_{12} +
\frac{11}{4}
\frac{C_A}{T} B_{12} ,
\label{A12AfterBLMSUN}
\\
\widetilde C_{12}
&=&
- \frac{3}{16}
  \frac{C_A}
       {T}
  (7C_A+11C_F)
  B_{12}
+ C_{12}
+ \frac{11}{4}
  \frac{C_A}{T}
  D_{12}
+ \frac{121}{16}
  \frac{C_A^2}{T^2}
  E_{12} ,
\label{C12AfterBLMSUN}
\\
Q^*
&=&
Q \exp
\Biggl[ \
   \frac{3}{2 T}
   B_{12}
   + \frac{9}{8 T^2}
   \left( \frac{11}{3} C_A
        - \frac{4}{3} T f
   \right)
   \left(B_{12}^2-E_{12}
   \right)
   \frac{\alpha_2(Q)}
        {\pi} \
\Biggr] ,
\label{QStarSUN}
\\
Q^{**}
&=&
Q \exp
\Biggl\{
\frac{3}{4 T}
\widetilde A_{12}^{-1}
\biggl[ \
    -\frac{1}{4} (5 C_A + 3 C_F) B_{12}
    + D_{12}
    + \frac{11}{2} \frac{C_A}{T} E_{12} \
\biggr] \
\Biggr\} .
\label{QStarStarSUN}
\end{eqnarray}
Notice the presence of $\alpha_2(Q)/\pi$ in the expression of $Q^*$.  In
general $Q^*$ will itself be a perturbative series in $\alpha_2(Q)/\pi$.
We have exponentiated the perturbative series, since physically, the
renormalization scale $Q^*$ should always be positive.  To the order
considered here, the scale for the coupling constant in Eq.
(\ref{QStarSUN}) is not well-defined, but can be chosen to be $Q$.  This
intrinsic uncertainty is similar to the $Q^{***}$ scale uncertainty of
the NNLO term in Eq. (\ref{Alpha1Alpha2AfterBLM}), and can only be
resolved by going to the next-higher order.

As an illustrative example, we quote the perturbative series of
$\alpha_{g_1}(Q)/\pi$ using dimensional regularization and the $\rm
\overline{MS}$ scheme with the renormalization scale fixed at $\mu=Q$:
\cite{LarinVermaseren}
\begin{eqnarray}
\frac{\alpha_{g_1}(Q)}
     {\pi}
&=&
\frac{\alpha_{\rm \overline{MS}}(Q)}
     {\pi}
+
\left(
\frac{\alpha_{\rm \overline{MS}}(Q)}
     {\pi}
\right)^2
\left[
\frac{23}{12} C_A - \frac{7}{8} C_F
-
\frac{1}
     {3}
f
\right]
\nonumber
\\
& &+
\left(
\frac{\alpha_{\rm \overline{MS}}(Q)}
     {\pi}
\right)^3
\Biggl\{
\left(
    \frac{5437}{648}
   -\frac{55}{18} \zeta_5
\right) C_A^2
+
\left(
   -\frac{1241}{432}
   +\frac{11}{9} \zeta_3
\right) C_A C_F
+
\frac{1}{32} C_F^2
\nonumber
\\
& &\hspace{2.5cm}
+
\left[
    \left(
         -\frac{3535}{1296}
         -\frac{1}{2} \zeta_3
         +\frac{5}{9} \zeta_5
    \right) C_A
+   \left(
          \frac{133}{864}
         +\frac{5}{18} \zeta_3
    \right) C_F
\right] f
\nonumber
\\
& &\hspace{2.5cm}
+
\frac{115}
     {648}
f^2
\Biggr\} .
\end{eqnarray}

The effective charge for the annihilation cross section has been computed
in the $\rm \overline{MS}$ scheme with the renormalization scale fixed at
$\mu=Q=\sqrt s$ \cite{GorishnyKataevLarin,SurguladzeSamuel}.  The
perturbative series for $\alpha_R(Q)/\pi$ is (using $T=1/2$ for the trace
normalization) \begin{eqnarray}
\frac{\alpha_R(Q)}
     {\pi}
&=&
\frac{\alpha_{\rm \overline{MS}}(Q)}
     {\pi}
+
\left(
\frac{\alpha_{\rm \overline{MS}}(Q)}
     {\pi}
\right)^2
\left[
  \left(
        \frac{41}{8}
       -\frac{11}{3} \zeta_3
  \right) C_A
-\frac{1}{8} C_F
+
\left(
- \frac{11}
       {12}
+
\frac{2}
     {3}
\zeta_3
\right) f
\right]
\nonumber
\\
& &+
\left(
\frac{\alpha_{\rm \overline{MS}}(Q)}
     {\pi}
\right)^3
\Biggl\{
  \left(
         \frac{90445}{2592}
        -\frac{2737}{108} \zeta_3
        -\frac{55}{18} \zeta_5
        -\frac{121}{432} \pi^2
  \right)
  C_A^2
\nonumber
\\
& & \hspace{2.5cm}
+ \left(
       -\frac{127}{48}
       -\frac{143}{12} \zeta_3
       +\frac{55}{3} \zeta_5
  \right)
  C_A C_F
-\frac{23}{32} C_F^2
\nonumber
\\
& &\hspace{2.5cm}
+
\biggl[
   \left(
        -\frac{970}{81}
        +\frac{224}{27} \zeta_3
        +\frac{5}{9} \zeta_5
        +\frac{11}{108} \pi^2
   \right)
   C_A
\nonumber
\\
& &\hspace{2.8cm}
+  \left(
        - \frac{29}{96}
        + \frac{19}{6} \zeta_3
        - \frac{10}{3} \zeta_5
   \right)
   C_F
\biggr]
f
\nonumber
\\
& &\hspace{2.5cm}
+
\left(
\frac{151}
     {162}
-
\frac{19}
     {27}
\zeta_3
-
\frac{1}
     {108}
\pi^2
\right)
f^2
\nonumber
\\
& &\hspace{2.5cm}
+
\left(
\frac{11}
     {144}
-
\frac{1}
     {6}
\zeta_3
\right)
\frac{d^{abc}d^{abc}}
     {C_F d(R)}
\frac{\left( \sum_f Q_f
      \right)^2}
     {\sum_f Q_f^2}
\Biggr\} .
\end{eqnarray}
The term containing $(\sum_f Q_f)^2/ \sum_f Q_f^2$ arises from
light-by-light diagrams.  The dimension of the quark representation is
$d(R)$, which usually is $N$ for $SU(N)$.  For QCD we have $d^{abc}
d^{abc} = 40/3 $.

The application of the NLO BLM formulas then leads to
\begin{eqnarray}
\frac{\alpha_{g_1}(Q)}{\pi}
&=&
\frac{\alpha_R(Q^*)}{\pi}
-
\frac{3}{4} C_F
\left(
   \frac{\alpha_R(Q^{**})}{\pi}
\right)^2
\nonumber
\\
& &
+
\left[
\frac{9}{16} C_F^2
-
\left(
\frac{11}
     {144}
-
\frac{1}
     {6}
\zeta_3
\right)
\frac{d^{abc}d^{abc}}
     {C_F N}
\frac{\left( \sum_f Q_f
      \right)^2}
     {\sum_f Q_f^2}
\right]
\left(
   \frac{\alpha_R(Q^{***})}{\pi}
\right)^3 ,
\\
Q^*
&=&
Q \exp
\left[
\frac{7}{4}
-
2 \zeta_3
+
\left(
\frac{11}{96}
+\frac{7}{3} \zeta_3
-2 \zeta_3^2
-\frac{\pi^2}{24}
\right)
\left(
   \frac{11}{3} C_A
  -\frac{2}{3} f
\right)
\frac{\alpha_R(Q)}{\pi}
\right],
\\
Q^{**}
&=&
Q \exp
\left[
 \frac{523}{216}
+\frac{28}{9} \zeta_3
-\frac{20}{3} \zeta_5
+
\left(
  -\frac{13}{54}
  +\frac{2}{9} \zeta_3
\right)
\frac{C_A}{C_F}
\right] .
\end{eqnarray}

(The scale $Q^{***}$ in the above expression can be chosen to be
$Q^{**}$.) Notice that aside from the light-by-light contributions, all
the $\zeta_3, \zeta_5$ and $\pi^2$ dependencies have been absorbed into
the renormalization scales $Q^*$ and $Q^{**}$.  Understandably, the
$\pi^2$ term should be absorbed into renormalization scale since it comes
from the analytical continuation of $R(Q)$ to the Euclidean region.

For the three flavor case, where we can neglect the light-by-light
contribution, the series remarkably simplifies to the CSR of Eq.
(\ref{AlphaG1AlphaRAfterBLMThreeFlavors}).  The form suggest that for the
general $SU(N)$ group the natural expansion parameter is $\widehat\alpha=
(3 C_F/4 \pi)\,\alpha$.  The use of $\widehat\alpha$ also makes explicit
that the same formula is valid for QCD and QED.  That is, in the limit
$N_C \to 0$ the perturbative coefficients in QCD coincide with the
perturbative coefficients of an Abelian analog of QCD.

\vskip1cm
\leftline{{\bf 3.
Commensurate Scale Relations and the Crewther Relation}}
\vskip.5cm

Broadhurst and Kataev have recently observed a number of interesting
relations between $\alpha_R(Q)$ and $\alpha_{g_1}(Q)$ (the ``Seven
Wonders") \cite{BroadhurstKataev}.  In particular, they have shown the
factorization of the beta function in the correction to the Crewther's
relation \cite{Crewther} which establishes a non-trivial connection
between the total $e^+e^-$ annihilation cross section and the polarized
Bjorken sum rule.  The simple form of our result Eq.
(\ref{AlphaG1AlphaRAfterBLMThreeFlavors}) also points to the existence of
a ``secret symmetry" between $\alpha_R(Q)$ and $\alpha_{g_1}(Q)$ which is
revealed after the application of the NLO BLM scale setting procedure.
In fact, as pointed out by Kataev and Broadhurst \cite{BroadhurstKataev},
in the conformally invariant limit, {\it i.e.}, for vanishing beta
functions, Crewther's relation becomes
\begin{equation}
(1+\widehat{\alpha}_R^{\rm eff})(1-\widehat{\alpha}_{g_1}^{\rm eff})=1.
\end {equation}
which is equivalent to our result in Eq.
(\ref{AlphaG1AlphaRAfterBLMThreeFlavors}).  Thus Eq.
(\ref{AlphaG1AlphaRAfterBLMThreeFlavors}) can be regarded as the
extension of the Crewther relation to non-conformally invariant gauge
theory.

The commensurate scale relation between $\alpha_{g_1}$ and $\alpha_R$
given by Eq. (14) implies that the radiative corrections to the
annihilation cross section and the Bjorken (or Gross-Llewellyn Smith) sum
rule cancel.  The relation between the physical cross sections can be
written in the forms:
\begin{equation}
{R_{e^+ e^-}(s)\over 3\sum e^2_q} ~
{\int^1_0 dx  g_1^p(x,Q^2)-g_1^n(x,Q^2) \over {1\over 3} g_A/g_V}
= 1 - \Delta \beta_0 \widehat a^3
\end{equation}
and
\begin{equation}
{R_{e^+ e^-}(s)\over 3\sum e^2_q} ~
{\int^1_0 dx  F_3^{\nu p}(x,Q^2) + F_3^{\bar\nu p}(x,Q^2)
\over 6} = 1 + \Delta \beta_0 \widehat a^3,
\end{equation}
provided that the annihilation energy in $R_{e^+ e^-}(s)$ and the
momentum transfer $Q$ appearing in the deep inelastic structure functions
are commensurate at NLO: $\sqrt s = Q^* = Q \exp [{7\over 4}- 2\zeta_3 +
({11\over 96} +{7\over 3} \zeta_3 - 2\zeta^2_3 - {\pi^2\over 24})\beta_0
\widehat a(Q)]$

The light-by-light correction to the CSR for the Bjorken sum rule
vanishes for three flavors.  The term $\Delta \beta_0 \widehat a^3$ with
$\Delta = \ell n\, (Q^*/Q^{**})$ is the third-order correction arising
from the difference between $Q^{**}$ and $Q^*$; in practice this
correction is negligible: for a typical value $\widehat a = \alpha_R(Q)/4
\pi = 0.14,$ $\Delta \beta_0 \widehat a^3 = 0.00071.$ Thus at the magic
energy $\sqrt s = Q^*$, the radiative corrections to the Bjorken and GLLS
sum rules almost precisely cancel the radiative corrections to the
annihilation cross section.  This allows a practical test and extension
of the Crewther relation to non-conformal QCD.

As an initial test of this CSR, we can compare the recent CCHS
measurement of the Gross-Llewellyn Smith sum rule
$1-\widehat\alpha_{F_3}={1 \over 6}\int^1_0 dx [F_3^{\nu p}(x,Q^2) +
F_3^{\bar\nu p}(x,Q^2)] = {1\over 3} ( 2.5 \pm 0.13 )$ at $Q^2 = 3$
GeV$^2$ and the parameterization of the annihilation data
\cite{MattSteven} $1 + \widehat\alpha_R = R_{e^+ e^-}(s)/3\sum e^2_q =
1.20.$ at the commensurate scale $\sqrt s = Q^*= 0.38\, Q = 0.66$ GeV.
The product is $(1 + \widehat\alpha_R)(1 -\widehat\alpha_{F_3})=1.00 \pm
0.04$, which is a highly nontrivial check of the theory at very low
physical scales.

\vskip1cm
\leftline{{\bf 4.
Other Applications of CSR}}
\vskip.5cm

As another example of a beyond-leading-order commensurate scale relation,
we shall express the effective charge for $\tau$ decay
$\alpha_\tau(M_\tau)/\pi$ in terms of the effective charge for $e^+ e^-$
annihilation $\alpha_R(Q)/\pi$.  The appropriate number of flavors in
this case is $f=3$, because $\tau$ decay occurs below the charm
threshold. [Incidentally, the light-by-light contribution in
$\alpha_R(Q)/\pi$ vanishes for the three flavor case.] The application of
the NLO BLM formulas leads to the following commensurate scale relation
\begin{eqnarray}
\frac{\alpha_\tau(M_\tau)}
     {\pi}
&=&
\frac{\alpha_R(Q^*)}
     {\pi} ,
\label{AlphaTauAlphaRAfterBLM}
\\
Q^*
&=&
M_\tau
\exp
\left[
   - \frac{19}{24}
   - \frac{169}{128}
   \frac{\alpha_R(M_\tau)}
        {\pi}
\right] .
\end{eqnarray}
Notice that all the $\zeta_3, \zeta_5$ and $\pi^2$ terms present in the
perturbative series of $\alpha_R(Q)/\pi$ and $\alpha_\tau(M_\tau)/\pi$
have disappeared when we related these two physical observables directly.
Notice also the vanishing NLO and NNLO coefficient in Eq.
(\ref{AlphaTauAlphaRAfterBLM}).  That is, up to the NNLO, the two
effective charge are simply related by a BLM scale shift.

Since the radiative corrections to the Bjorken sum rule are identical to
those of the Gross-Llewellyn-Smith sum rule---up to small corrections of
order $\alpha^3_s(Q^2)$, a basic test of QCD can be made by considering
the ratio of the Gross-Llewellyn-Smith and Bjorken sum rules:
\begin{equation}
R_{GLLS/Bj}(Q^2,\epsilon) =  {
\frac{1}{6}
\int^1_\epsilon dx
\left[ F^{\nu p}_3(x,Q^2)+F^{\bar \nu p}_3(x,Q^2) \right]\over
3 \left|  \frac{g_V}{g_A}
  \right|
\int^1_\epsilon dx \left[ g_1^p(x,Q^2)-g_1^n(x,Q^2) \right]}.
\end{equation}
If the Regge behavior of the two sum rules is similar, the empirical
extrapolation to $\epsilon \rightarrow 0$ should be relatively free of
systematic error. Moreover, PQCD predicts
\begin{equation}
 R_{GLLS/Bj}(Q^2, \epsilon \to 0) = 1
+ {\cal O}\left(\alpha^3_s(Q)\right) +
{\cal O}\left( \Lambda_{QCD}^2\over Q^2\right) \ ,
\end{equation}
{\it i.e.}, hard relativistic corrections to the ratio of the sum rules
only enter at three loops.  Thus measurements of the {\it ratio} of the
sum rules could provide a remarkably complication-free test of QCD - any
significant deviation from $R_{GLLS/Bj}(Q^2,\epsilon \to 0)=1$ must be
due to higher twist effects which should vanish rapidly with increasing
$Q^2.$

After scale-fixing, the ratio of hadronic to leptonic decay rates for the
$\Upsilon$ has the form \cite{BrodskyLepageMackenzie}
\begin{eqnarray}
{\Gamma(\Upsilon  \to {\rm hadrons})\over
\Gamma(\Upsilon  \to \mu^+ \mu^-)}
&=&
{10 (\pi^2-9) \over 81 \pi e_b^2}\
{\alpha^3_{\overline{\rm MS}}(0.157 M_\Upsilon)\over \alpha_{\rm QED}^2}
\left[1 - 14.0(5)\, {\alpha_{\overline{\rm MS}}\over \pi}+\cdots \right]
\\
&=&
{10 (\pi^2-9) \over 81 \pi e_b^2}\
  {\alpha^3_V(0.363 M_\Upsilon)\over \alpha_{\rm QED}^2}
  \left[1 - 8.0(5)\, {\alpha_V\over \pi} + \cdots \right] .
\end{eqnarray}
Thus, as is the case of positronium decay, the next to leading
coefficient is very large, and perturbation theory is not likely to be
reliable for this observable.  On the other hand, the commensurate scales
for the second moment of the non-singlet structure function $M_2$ and the
effective charges in the Bjorken Sum Rule (and the Gross-Llewellyn-Smith
Sum Rule) are not far from the physical value $Q$ when expressed in
$\alpha_V$ scheme.  At large $n$ the commensurate scale for $M_n$ is
proportional to $1/\sqrt n$, reflecting the fact that the available
phase-space for parton emission decreases as $n$ increases.  In
multiple-scale cases, the commensurate scale can depend on all of the
physical invariants.  For example, the scale controlling the evolution
equation for the non-singlet structure function depends on $x_{Bj}$ as
well as $Q$ \cite{Wong}.  In the case of inclusive reactions which
factorize at leading twist, each structure function, fragmentation
function, and subprocess cross section can have its own scale.

After one fixes the renormalization scale $\mu$ to the BLM value, it is
still useful to compute the logarithmic derivative of the truncated
perturbative prediction $d \ell n\, \rho_N/ d \ell n \mu$ for a physical
observable $\rho$ at the BLM-determined scale.  If this derivative is
large, or equivalently, if the BLM and PMS \cite{Stevenson} scales
strongly differ, then one knows that the truncated perturbative expansion
cannot be numerically reliable, since the entire series is independent of
$\mu.$ Note that this is a necessary condition for a reliable series, not
a sufficient one, as evidenced by the large coefficients in the
positronium and quarkonium decay widths which appear when the scales are
set correctly.  In the case of the two and three jet decay fractions in
$e^+ e^-$ annihilation, the BLM and PMS scales strongly differ at low
values of the jet discriminant $y.$ Thus, by using this criterion, we
establish that perturbation theory must fail in the small $y$ regime,
requiring careful resummation of the $\alpha_s \ell n\, y$ series. (A
more detailed discussion of the sensitivity of the jet fractions to scale
choice and jet clustering schemes is given in Ref. \cite{BurrowsMasuda}.)

However, if we restrict the analysis to jets with invariant mass ${\cal
M} < \sqrt{y s},$ with $0.14 > y > 0.05$, then we have an ideal
situation, since both the PMS and FAC scales nearly coincide with the BLM
scale when one computes jet ratios in the $\overline{\rm MS}$ scheme {\it
i.e.}, the renormalization scale dependence in this case is minimal at
the BLM scale, and the computed NLO (next-to-leading order) coefficient
is nearly zero.  In fact, Kramer and Lampe \cite{KramerLampe} find that
the BLM scale and the NLO PQCD predictions give a consistent description
of the LEP 2-jet and 3-jet data for $0.14 > y > 0.05$ at the $Z.$
Neglecting possible uncertainties due to hadronization effects, this
allows a determination of $\alpha_s$ with remarkably small
error:\cite{KramerLampe}
$\alpha_{\overline{\rm MS}} (M_z)=0.107\pm 0.003,$ which
corresponds to $\Lambda^{(5)}_{\overline{\rm MS}} = 100 \pm 20 $ MeV.
It is clear that reanalyses of the SLD and LEP data will need
to be done with BLM scale breaking.

\vskip1cm
\leftline{{\bf 5. Conclusions}}
\vskip.5cm

The commensurate scale relations open up additional possibilities for
testing QCD.  One can compare two observables by checking that their
effective charges agree both in normalization and in their scale
dependence.  The ratio of commensurate scales $\lambda_{A/B}$ is fixed
uniquely: it ensures that both observables $A$ and $B$ pass through heavy
quark thresholds at precisely the same physical point.  The same
procedure can be applied to multi-scale problems; in general, the
commensurate scales $Q^*, Q^{**}$, etc. will depend on all of the
available scales.

Calculations are often performed most advantageously in $\overline{\rm
MS}$ scheme, but all reference to such theoretically constructed schemes
have to vanish when comparisons are made between observables.  We
emphasize that any consistent renormalization scheme, with any arbitrary
choice of renormalization scale $\mu,$ can be used in the intermediate
stages of analysis.  The final result, the commensurate scale relation
between observables, is guaranteed to be independent of the choice of
intermediate renormalization scheme since the BLM procedure satisfies the
generalized renormalization group properties of Peterman and Stuckelberg.
An important computational advantage is that one only needs to compute
the $f$-dependence of the higher order terms in order to specify the
lower order scales in the commensurate scale relations.  In many cases,
the series coefficients in the commensurate scale relations can be
determined from the corresponding Abelian theory; {\it i.e.} $N_C \to 0.$

The BLM method and the commensurate scale relations presented here can be
applied to the whole range of QCD and standard model processes, making
the tests of theory much more sensitive.  The method should also improve
precision tests of electroweak, supersymmetry and other non-Abelian
theories.  One of the most interesting and important areas of application
of commensurate scale relations will be to the hadronic corrections to
exclusive and inclusive weak decays of heavy quark systems, since the
scale ambiguity in the QCD radiative corrections is at present often the
largest component in the theoretical error entering electroweak
phenomenology.

We have also presented in this talk a number of other commensurate scale
relations using the extension of the BLM method to the next-to-leading
order.  We have shown that in each case the application of the NLO BLM
formulas to relate known physical observables in QCD leads to results
with surprising elegance and simplicity.  The commensurate scale
relations for some of the observables ($\alpha_R,
\alpha_\tau, \alpha_{g_1}$ and $\alpha_{F_3}$)
are universal in the sense that the coefficients of $\widehat \alpha_s$
are independent of color; in fact, they are the same as those for Abelian
gauge theory.  Thus much information on the structure of the non-Abelian
commensurate scale relations can be obtained from much simpler Abelian
analogs.  In fact, in the examples we have discussed here, the
non-Abelian nature of gauge theory is reflected in the $\beta$-function
coefficients and the choice of second-order scale $Q^{**}.$

Because they relate observables to observables, the commensurate scale
relations are convention-independent; {\it i.e.}, independent of the
normalization conventions used to define the color $SU(N)$ matrices, etc.
Since the ambiguities due to scale and scheme choice have been
eliminated, one can ask fundamental questions concerning the nature of
the QCD perturbative expansions, {\it e.g.}, whether the series is
convergent or asymptotic, due to renormalons, etc.\cite{Mueller}.

The commensurate scale relations between observables can be tested at
quite low momentum transfers, even where PQCD relationships would be
expected to break down.  It is possible that some of the higher twist
contributions common to the two observables are also correctly
represented by the commensurate scale relations.  In contrast, expansions
of any observable in $\alpha_{\overline{\rm MS}}\,(Q)$ must break down at
low momentum transfer since $\alpha_{\overline{\rm MS}}\,(Q)$ becomes
singular at $Q=\Lambda_{\overline{\rm MS}}.$ (For example, in the 't
Hooft scheme where the higher order $\beta_n=0$ for $n=2,3,...$ ,
$\alpha_{\overline{\rm MS}}(Q)$ has a simple pole at
$Q=\Lambda_{\overline{\rm MS}}.$) The commensurate scale relations allow
tests of QCD in terms of finite effective charges without explicit
reference to singular schemes such as $\overline{\rm MS}.$
The coefficients in the CSR are identical to the coefficients
in a conformal theory where renormalons do not appear.  It is thus
reasonable to expect that the series expansions appearing in the CSR are
convergent when one relates finite observables to each other.  Thus
commensurate scale relations between observables allow tests of
perturbative QCD with higher and higher precision as the perturbative
expansion grows.

A natural procedure for developing a precision QCD phenomenology is to
choose one effective charge as the canonical definition of the QCD
coupling, and then predict all other observables in terms of this
canonical measure.  Ideally, the heavy quark effective charge
$\alpha_V(Q^2)$ could serve this central role since it can be determined
from both the quarkonium spectrum and from lattice gauge theory.
However, it will be necessary to compute the relation of the heavy quark
potential to other schemes through three loops.  At present, the most
precisely theoretically and empirically known effective coupling is
$\alpha_R(Q^2),$ as determined from the annihilation cross section; thus
it is natural to use it as the standard definition.
Alternatively, one can follow historical convention and continue to use
the $\overline {\rm MS}$ scheme as an intermediary between observables.
For definiteness, let us consider a 't Hooft scheme
with $\Lambda=\Lambda_{\overline{\rm MS}}$ having all $\beta_n = 0$
beyond $n=1.$ The commensurate scale relations such as Eq.
(\ref{QStarSUN}) and (\ref{QStarStarSUN}) then unambiguously specify all
of the scales $Q^*, Q^{**}$, etc. required to relate
$\alpha_{\overline{\rm MS}}$ to the observables.  The intrinsic QCD scale
will then be unambiguously encoded as $\Lambda_{\overline{\rm MS}}$.
However, as we have emphasized,there is an intrinsic disadvantage in
using $\alpha_{\overline{\rm MS}}(Q)$ as an expansion parameter: the
function $\alpha_{\overline{\rm MS}}(Q)$ has a simple pole at
$Q=\Lambda_{\overline{\rm MS}}$, whereas observables are by definition
finite.

\vskip1cm
\leftline{{\bf 6.  Acknowledgments}}
\vskip.5cm

We wish to thank Andrei Kataev for helpful discussions and to Bennie Ward
for organizing this interesting conference.  This work is supported in
part by the Department of Energy, contract DE--AC03--76SF00515 and
contract DE--FG02--93ER--40762.


\begin{references}
\bibitem{BrodskyLuBig}
S. J. Brodsky and H. J. Lu, SLAC-PUB-6481 (1994).
\bibitem{Grunberg}
G. Grunberg,
\sl Phys. Lett. \bf B95 \rm (1980) 70;
\sl Phys. Lett. \bf B110\rm (1982) 501;
\sl Phys. Rev. \bf D29 \rm (1984) 2315.
\bibitem{DharGupta}
A. Dhar and V. Gupta,
\sl Phys. Rev. \bf D29 \rm (1984) 2822.
\bibitem{GuptaShirkovTarasov}
V. Gupta, D.V. Shirkov and O.V. Tarasov,
\sl Int. J. Mod. Phys. \bf A6 \rm (1991) 3381.
\bibitem{LarinVermaseren}
S. A. Larin and J.A.M. Vermaseren,
\sl Phys. Lett. \bf B259 \rm (1991) 345.
\bibitem{EllisKarliner}
J. Ellis and M. Karliner,
\sl Phys. Lett. \bf B313 \rm (1993) 131.
\bibitem{BrodskyLuStueckelbergPeterman}
S. J. Brodsky and H. J. Lu,
\sl Phys. Rev. \bf D48 \rm (1993) 3310.
\bibitem{StueckelbergPeterman}
E. C. G. St\"uckelberg and A. Peterman,
\sl Helv. Phys. Acta \bf 26 \rm (1953) 499,
A. Peterman,
\sl Phys. Rept. \bf 53C \rm (1979) 157.
\bibitem{BrodskyLuSelfconsistency}
S. J. Brodsky and H. J. Lu,
SLAC-PUB-6000 (1993).
\bibitem{GellMannLow}
M. Gell-Mann and F. E. Low,
\sl Phys. Rev. \bf 95 \rm (1954) 1300.
\bibitem{BogoliubovShirkov}
N. N. Bogoliubov and D. V. Shirkov,
\sl Dokl. Akad. Nauk \bf 103 \rm (1955) 391.
\bibitem{Leipzig}
A preliminary report of this work appears in
S. J. Brodsky and H. J. Lu,
Invited talk presented at {\it Leipzig Workshop on Quantum Field
Theoretical Aspects of High Energy Physics}, Bad Frankenhausen,
Germany, September 20-24, 1993. SLAC-PUB-6389.
\bibitem{BrodskyLepageMackenzie}
S. J. Brodsky, G. P. Lepage and P. B. Mackenzie,
\sl Phys. Rev. \bf D28 \rm (1983) 228.
\bibitem{LepageMackenzie}
G. P. Lepage, P. B. Mackenzie,
\sl Phys. Rev. \bf D48 \rm  (1993) 2250.
\bibitem{Davies}
C. T. H. Davies, K. Hornbostel, G. P. Lepage,
A. Lidsey, J. Shigemitsu, and
J. Sloan,  Cornell preprint OHSTPY-HEP-T-94-013 (1994).
\bibitem{HeavyQuarkPotential}
W. Fischler,
\sl Nucl. Phys. \bf B129 \rm (1977) 157,
A. Billoire,
\sl Phys. Lett. \bf B92 \rm (1980) 343,
W.~Buchmuller, G. Grunberg and S.H.H. Tye,
\sl Phys. Rev. Lett. \bf 45 \rm (1980) 103;
{\bf 45} \rm (1980) 587(E).
\bibitem{KramerLampe}
G. Kramer and B. Lampe,
\sl Zeit. Phys. \bf A339 \rm  (1991) 189.
\bibitem{IngelmanRathsman}
G. Ingelman and J. Rathsman,
preprint TSL-ISV-94-0096.
\bibitem{VoloshinSmith}
B. H. Smith and M. B. Voloshin,
preprint TPI-MINN-94-16-T.
\bibitem{Sirlin}
A. Sirlin,
preprint NYU-TH-94-08-01.
\bibitem{GorishnyKataevLarin}
S. G. Gorishny, A. L. Kataev and S. A. Larin,
\sl Phys. Lett. \bf B259 \rm (1991) 144.
\bibitem{SurguladzeSamuel}
L. R. Surguladze and M. A. Samuel,
\sl Phys. Rev. Lett. \bf 66 \rm (1991) 560,
{\bf 66} \rm (1991) 2416(E).
\bibitem{BroadhurstKataev}
D. J. Broadhurst and A. L. Kataev,
\sl Phys. Lett. \bf B315 \rm (1993) 179.
\bibitem{Crewther}
R. J. Crewther,
\sl Phys. Rev. Lett. \bf 28 \rm (1972) 1421.
\bibitem{MattSteven}
A. C. Mattingly and P. M. Stevenson,
\sl Phys. Rev. \bf D49 \rm (1994) 437.
\bibitem{Wong}
W. K. Wong, {\it et al.}, (in preparation).
\bibitem{Stevenson}
P. M. Stevenson,
\sl Phys. Lett. \bf B100 \rm (1981) 61;
\sl Phys. Rev. \bf D23 \rm (1981) 2916;
\sl Nucl. Phys. \bf B203 \rm (1982) 472;
\sl Nucl. Phys. \bf B231 \rm (1984) 65.
\bibitem{BurrowsMasuda}
P. N. Burrows and H. Masuda,
preprint SLAC-PUB-6394 (1993).
\bibitem{Mueller}
A. H. Mueller,
\sl Phys. Lett. \bf B308 \rm (1993) 355.
\end{references}
\end{document}